\documentclass[preprint,amsmath,amssymb,showpacs, aps,longbibliography]{revtex4-1}

\usepackage[dvips]{graphicx}

\begin{document}

\title{Macroscopically constrained Wang-Landau method for systems with multiple order parameters and its application to drawing complex phase diagrams}




\author{C.~H. Chan$^{1}$}
\email{seahoi2001@gmail.com }

\author{G. Brown$^{1,2}$}
\email{gbrown@fsu.edu}

\author{P.~A. Rikvold$^{1}$}
\email{prikvold@fsu.edu}

\affiliation{$^1$ Department of Physics, Florida State University, Tallahassee, Florida 32306-4350, USA\\
$^{2}$ Division of Science and Math, Tallahassee Community College, Tallahassee, Florida 32304, USA}

\date{\today }

\begin{abstract}
A generalized approach to Wang-Landau simulations, macroscopically constrained
Wang-Landau, is proposed to simulate the density of states of a system with multiple
macroscopic order parameters. 
The method breaks a multidimensional random-walk
process in phase space into many separate, one-dimensional random-walk processes in well-defined
subspaces. Each of these random walks is constrained to a different set of values of the
macroscopic order parameters.
When the multi-variable density of states is obtained for one set of
values of field-like model parameters,
the density of states for any other values of these parameters can be obtained by
a simple transformation of the total system energy. All thermodynamic quantities
of the system can then be rapidly calculated at any point in the phase diagram.
We demonstrate how to use the multi-variable density of states to draw the
phase diagram, as well as order-parameter probability distributions at specific phase points,
for a model spin-crossover material:
an  antiferromagnetic Ising model with ferromagnetic long-range interactions.
The field-like parameters in this model are an effective magnetic field and the strength of the
long-range interaction.
\end{abstract}


\maketitle

\section{introduction}\label{sec:Introduction}

Classical spin models have found wide application to a vast array of problems in
many branches of physics and other sciences.
This is due to the relative simplicity of such models and
the fact that the different spin states can be given many interpretations
besides that of magnetic spins, including different kinds of atoms or molecules
(``lattice-gas models"), opinions, biological species in an ecosystem, etc.
A few examples of these diverse applications are magnetic materials \cite{RICH94},
high-energy physics \cite{CASE03}, astrophysics \cite{HASN13},
electrochemistry \cite{BROW99A},
polymer science \cite{LUO03}, network reliability problems \cite{PhysRevE.94.042125},
and economics \cite{SORN14}.
The archetypal member of this class of
models is the $S=1/2$ ferromagnetic Ising model of binary spin variables placed at the
sites of a lattice or a more general network and interacting via a simple Hamiltonian. Since its
introduction for a one-dimensional system almost a century ago \cite{Ising_1925},
this model has been joined by many antiferromagnetic or ferromagnetic
generalizations to higher spatial dimensions,
multiple local states and/or multidimensional order-parameter spaces,
such as the $S=1$ or three-state Blume-Capel \cite{PhysRev.141.517,CAPEL1966966}
and Blume-Emery-Griffiths models \cite{PhysRevA.4.1071}.
Although the order parameter for the $S=1/2$ square-lattice
Ising model in zero field has been obtained exactly \cite{PhysRev.65.117,YANG52},
solution of this class of models under general conditions, including in nonzero field, is known to be NP hard \cite{BARA82}.
As a consequence, much effort has been applied to
developing accurate approximate and numerical solutions, including
mean-field approximations \cite{BRAG34,BRAG35,AGRA06},
series expansions \cite{DOMB74},
numerical transfer-matrix calculations \cite{NIGH90},
and a variety of Monte Carlo (MC) methods \cite{Landau_simulation_book}.
Although many ingenious algorithms have been introduced, development of improved
numerical methods to study equilibrium and nonequilibrium aspects of classical spin systems
remains an active research area.

In this paper we present a generalization of the Wang-Landau (WL) MC method for calculating
densities of states (DOS)
\cite{PhysRevLett.86.2050,PhysRevE.64.056101} to systems with multiple order parameters.
Development of the method was inspired by applications
to a class of molecular crystals known as spin-crossover materials \cite{HALC13},
some of which can have competing antiferromagnetic-like short-range and
ferromagnetic-like long-range interactions.
A discrete-spin model of such a system was recently studied for a few
values of two field-like model parameters (an effective external field and the strength of
the long-range interaction)
by a computationally intensive Metropolis importance-sampling MC
method \cite{PhysRevB.93.064109}. In order to obtain results for a wide range of
model parameters with a
manageable computational effort, a simulation method is needed that
can produce three-dimensional
DOS, $g(E,M,M_s)$, where $E$ is the total system energy,
$M$ can be interpreted as a total system magnetization,
and $M_s$ as a staggered magnetization \cite{Endnote1}.
The method can also in principle deal with higher-dimensional order-parameter spaces.
The original time-consuming WL random-walk simulation in multiple-dimensional phase space is broken down into many stages. In each stage,
many independent WL simulations perform one-dimensional walks, each with different constrained macroscopic parameters.
For the lattice model we consider in this paper, exact combinatorial calculations can be applied to simplify the process, so that only one stage of simulation in $E$ is required. As the simulations are run independently, no special skills in parallel programming are required.
From the one-dimensional random-walk WL simulations in $E$,
performed separately over a grid in the order-parameter space
at one single set of model parameters, the method can produce DOS for any
value of model parameters and temperatures, using a simple transformation of the
total system energy.
This contrasts with both importance-sampling and the original WL MC methods, in which
separate simulations must be performed for each set of model parameters of interest.
From the resulting multidimensional DOS, properties such as phase diagrams, free-energy landscapes, and
joint and marginal order-parameter distributions can be simply obtained.

Another advantage of this method is the ability to use symmetries in the order-parameter space to reduce the
number of simulations needed. (For instance, in the case of the spin-crossover model, such
symmetry considerations lead to an additional eight-fold reduction in the computational work.)
For concreteness, the details of the method will be demonstrated here in the context of the
model spin-crossover material of Ref.\ \cite{PhysRevB.93.064109}.
Further results for several sets of model parameters of experimental interest
will be described in forthcoming papers \cite{CHAN17conference,CHAN17}.

The remainder of this paper is organized as follows. In Sec.~\ref{sec:Hamiltonian_Ising_ASFL} we introduce the model spin-crossover material Hamiltonian that inspired the method, and which we will use to illustrate the application of our algorithm.
In Sec.~\ref{sec:current_methods} we first summarize the relevant basics of the WL algorithm, and then we discuss several ways of finding the joint DOS, $g(E,M)$, previously introduced in the literature, pointing out their weaknesses when applied to the situation studied here. 
In Sec.~\ref{sec:our_WL} we discuss the macroscopically constrained WL algorithm in detail. Some calculations and symmetry considerations are discussed in Appendices A and B, and the detailed implementation of the method is described in Appendix C.
In Sec.~\ref{sec:application} we
give numerical results for the model spin-crossover material, and we demonstrate how to use $g(E,M,M_{s})$ to draw and investigate its phase diagram.
A discussion of methods to extend the system size
is given in Sec.~\ref{sec:generalizion big system}, and
conclusions and a brief discussion of future work 
are given in Sec.~\ref{sec:conclusion}.

\section{Two-dimensional Ising model with Antiferromagnetic Short-range and Ferromagnetic Long-range interactions (2D Ising-ASFL model)}\label{sec:Hamiltonian_Ising_ASFL}
To demonstrate the details and performance of the proposed method, and for its comparison
with other methods, we will use a pseudo-spin model of a spin-crossover material
with short-range antiferromagnetic-like and long-range ferromagnetic-like interactions,
which was previously introduced and studied by Metropolis importance-sampling MC in Refs.\
\cite{BROWN201420,PhysRevB.93.064109}.
It is defined by the Hamiltonian,
\begin{equation}\label{def_Hamiltonian_2D_Ising-ASFL}
{\mathcal H}= J \sum_{\langle i,j \rangle} s_i s_j - \frac{A}{2N} M^{2}  - HM \;.
\end{equation}
The local variables are $s_i=\pm 1$, and $M = \sum_i s_i$ is the corresponding
global ``magnetization."
The explicit sum runs over all nearest-neighbor pairs on an
$N=L \times L$ square lattice with periodic boundary conditions, and $J>0$ makes the
local interactions antiferromagnetic. The second term models long-range,
mean-field like ferromagnetic interactions of strength $A\geq 0$.
In the third term, $H$ is the applied field (actually an effective field in the spin-crossover model \cite{WAJN71,PhysRevB.77.014105,PhysRevB.84.054433,PhysRevB.93.064109}),
which breaks the symmetry between positive and negative $M$.
$A$ and $H$ are the model's two field-like parameters.
Throughout this paper, we will use the notation $E$ to represent
the total energy obtained through this Hamiltonian,
including the contributions from the terms proportional to $A$ and $H$.
$E$, $A$, and $H$ will be given in units of $J$, and temperatures in units of
$J/k_{\rm B}$ with $k_{\rm B}$ being Boltzmann's constant.

While this Hamiltonian
includes only the ferromagnetic order parameter $M$, we note that both linear and nonlinear
terms (in this case $M^2$) are included. Linear and/or nonlinear terms in the staggered magnetization $M_s$ (see below) could also
be added as needed to model other particular systems. However, the form given here is sufficient to
demonstrate and validate the algorithm.

\section{Basic Wang-Landau and current methods to find joint densities of states}\label{sec:current_methods}
In this section, we review some relevant basics of the WL method,
followed by some current methods to obtain joint DOS,
and we  explain why these methods are not appropriate to obtain $g(E,M,M_{s})$ for the
system defined by the Hamiltonian (\ref{def_Hamiltonian_2D_Ising-ASFL}).
These are mostly WL based methods.

 \subsection{Basic Wang-Landau Monte Carlo method}
The WL method is a restricted random-walk method for finding  the DOS of a system.
Its idea is based on the observation that if one imposes
an acceptance probability for proposed energy transitions in the random walk
which is proportional to
the {\it reciprocal\/} of the DOS, then the system will spend roughly equal times in all
different energy states. If a histogram $H(E)$ is used to record the number of times that the
walker visits each energy state, a `flat' histogram will eventually be generated.
The whole random walk process is divided into many sweeps. In each sweep, the estimated DOS is multiplied by
a modification factor, $f>1$. The sweep is finished when a `flat' histogram is obtained.
Then, the next sweep starts with a smaller value of $f$. The whole process ends when $\ln f<10^{-8}$.
To speed up the simulation,
a wide energy spectrum may be divided into separate
energy windows, with separate simulations performed in each window.
The partitioned windows may be uniform \cite{PhysRevLett.86.2050,PhysRevE.64.056101} or nonuniform \cite{PhysRevE.78.055701}.
As errors are generated near the window boundaries
\cite{PhysRevE.67.067102,PhysRevE.64.056101} whenever proposed moves outside the
window are rejected,
neighboring energy windows should overlap by a certain fraction, and
the estimated $g(E)$ obtained in neighboring windows should be joined at the point
where their slopes with respect to $E$ are closest,
so that a smooth $g(E)$ over the whole energy spectrum can be obtained.
In complex systems with rugged energy landscapes, states that lie in the same energy window may sometimes be connected only by paths that go via a different window.
Therefore, the Replica Exchange Wang-Landau (REWL) scheme \cite{PhysRevLett.110.210603,WaiReplica,PhysRevE.90.023302} was proposed to ensure that all microstates are
visited. The scheme allows two walkers that both have
energies within the overlap region of adjacent windows
to exchange their microstates with a certain probability, so that ergodicity is preserved.
REWL is performed in parallel
\cite{Yin20121568}.
Errors and convergence of WL have also been studied
\cite{PhysRevE.72.025701,Lee200636,PhysRevE.75.046701,content/aip/journal/jcp/127/18/10.1063/1.2803061,PhysRevE.84.065702,PhysRevE.85.010102}.
It was found that the statistical errors in $\ln g(E)$ are proportional to $\sqrt{\ln f}$ \cite{PhysRevE.72.025701}, and the fluctuations in the histogram
are proportional to $1/\sqrt{\ln f}$ \cite{Lee200636}.
The accuracy of $g(E)$ may be increased by using the $1/t$ algorithm \cite{PhysRevE.75.046701,PhysRevE.78.067701,content/aip/journal/jcp/127/18/10.1063/1.2803061}. A mathematical generalization of the WL algorithm is the Stochastic Approximation Monte Carlo (SAMC) algorithm \cite{doi:10.1198/016214506000001202,Liang2009581}, which has also involved the concept used in the $1/t$ algorithm. Recently, Junghans \textit{et al.} have demonstrated
that WL, statistical temperature molecular dynamics, and metadynamics are
equivalent under consistent initial conditions and update rules \cite{doi:10.1021/ct500077d}.

\subsection{Wang-Landau with multi-dimensional random walk in phase space}\label{sec_2D_WL}
The basic WL method for finding the {\it joint\/} DOS of a system, $g(E,V_{1})$, is to perform
 a two-dimensional random walk in the $(E,V_{1})$
space \cite{PhysRevE.64.056101,KWAK201280,PhysRevE.73.036702,PhysRevE.75.061108,2DWLAlex}.
However, this approach is quite slow. To speed up the simulation, the system could be divided into multiple energy windows, with each window containing all the 
compatible $V_{1}$, using
replica-exchange to ensure that all the microstates are accessible in each energy window \cite{2DWLAlex}.
However,  if the {\it joint\/} DOS contains one more variable, 
  like the $g(E,M,M_{s})$ we want to obtain here, 
  the simulation will again become slow. Here we performed several crude tests on the antiferromagnetic Ising model with different system sizes using this method.

  We first adopt a strict `flatness' criterion similar to the original WL papers \cite{PhysRevLett.86.2050,PhysRevE.64.056101}, which considers a histogram `flat' if for every state, the deviation in the histogram $H(E,M,M_{s})$ is less than $20\%$ from the average histogram.
 Using parallel programming with the energy spectrum divided into 5 energy windows, with each window assigned 5 random walkers and 1 core, a $6\times 6$ Ising system takes 23 min to finish the simulation, while an $8\times 8$ Ising system takes 972 min ($\sim$ 16 hours). This approach obviously does not scale well with system size.

   A relaxed `flatness' criterion \cite{BROWN201428} considers a histogram `flat' if the root-mean-square of the deviation from the average histogram is less then $20\%$, i.e.,
\begin{equation}\label{def_release_deviation1}
\sqrt{\frac{\sum_{E,M,M_{s}} |H(E,M,M_{s}) - H_{\rm average}|^{2}}{N_{E,M,M_{s}} }} < 20\% \; ,
\end{equation}
where $N_{E,M,M_{s}}$ is the number of accessible  $(E,M,M_{s})$ in the window. Using this criterion significantly improves the convergence times, but it does not improve the scaling behavior of this approach (see Table~\ref{table_speed}). A very rough  estimate for the time it would take for an $32 \times 32$ system to finish is $20000s \times 5^{10} \sim 6000$ years.

\begin{table}
\caption{Crude tests for the computational time to obtain $g(E,M,M_{s})$, using WL
with multi-dimensional random walk in phase space, with energy spectrum partitioned into windows plus replica-exchange \cite{2DWLAlex}. Parallel programming was performed with each window assigned 1 core and 5 walkers. The histogram is considered to be `flat' if all the walkers satisfy the root-mean-square `flatness' criterion \cite{BROWN201428} (Eq.~(\ref{def_release_deviation1})). The time recorded includes the time for initializing the systems.  The asterisks mark the cases in which too many windows were used for a small system. In general, changing the system size from $L$ to $L+2$ causes the simulation time to increase by a factor of 6 to 10.}
\centering
\begin{tabular}{llll}
\hline
\hline
$L\times L$ \ \ \  & 3 windows \ \  & 5 windows \ \  & 7 windows \\
\hline
$6\times 6$  \ \ \     &   $38$ s  \ \     &   $33$ s \ \   &   *$56$ s    \\
$8\times 8$  \ \ \     &   $636$ s   \ \   &   $430$ s  \ \    &   *$1128$ s      \\
$10\times 10$  \ \ \    &   $6572$ s  \ \    &   $4541$ s  \ \    &   $2720$ s      \\
$12\times 12$  \ \ \     &   $52728$ s  \ \    &   $29312$ s  \ \    &   $22052$ s      \\
\hline
\end{tabular}\label{table_speed}
\end{table}

\subsection{Two-stage method}\label{sec_1D_record_2D}

Another method to obtain a joint DOS breaks the simulation into two stages of random walk
\cite{content/aip/journal/jcp/130/21/10.1063/1.3148186,PhysRevE.87.012706,Kalyan2016}.
In the first stage, a normal WL process is carried out and $g(E)$ is obtained.
Then, a second stage of random walk is performed with an acceptance probability of $1/g(E)$, but only $H(E,M,M_{s})$ is updated, i.e. $g(E)$ remains unchanged. The joint DOS
is then obtained from
\begin{equation}\label{def_1D_to_2D_DOS}
g(E,M,M_{s})=\frac{ H(E,M,M_{s}) }{ \sum_{M,M_{s}}H(E,M,M_{s}) } g(E) ~.
\end{equation}
We partition the first-stage WL process into 5 energy windows with replica exchange. Each window has 5 walkers, and a separate core is assigned to each window. The strict `flatness' criterion requiring that every state $(E,M,M_{s})$ does not deviate more than $20\%$ from the average histogram is adopted \cite{PhysRevLett.86.2050,PhysRevE.64.056101}. In the second stage of random walk, each core is assigned a random walker in a random initial state, which can walk through the whole phase space. 
We set the number of time steps spent on stage 2 to be
10 times of that spent on stage 1. This method is much faster than the previous method:
an $8\times 8$ system can be finished in a few seconds, and even $12\times 12$ can be finished in 170 seconds if the walkers do not get `stuck' \cite{Yin20121568,PhysRevE.88.053302,PhysRevE.92.023306}.

\begin{figure}
\includegraphics[width=0.5\textwidth]{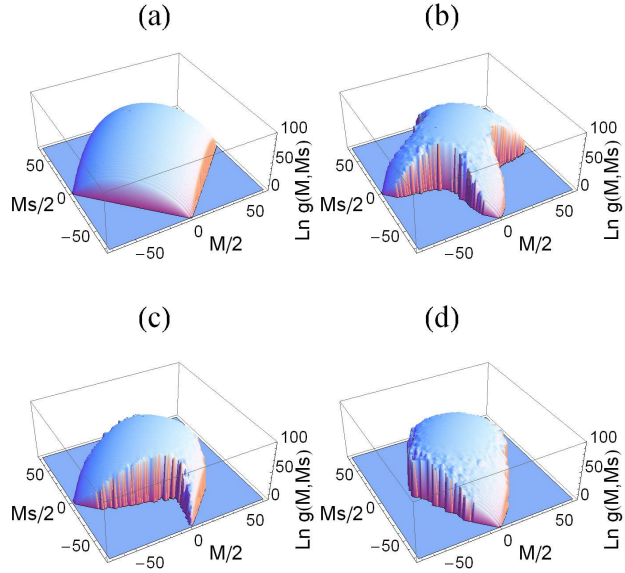}
\caption{(Color online) The two-dimensional joint DOS, $g(M,M_{s})$,  for $L=12$,
obtained by (a) exact combinatorial calculation as
described in Appendix~\ref{sec:exact_combinatorial}, and by the two-stage method, 
using different values of $H$ and $A$,
(b) $(H,A)=(0,0)$, (c) $(H,A)=(3,0)$, and (d) $(H,A)=(0,7)$. The $g(M,M_{s})$ should be independent of $H$ and $A$. The deviations from (a) of the results shown in
(b) to (d) imply that the two-stage method does not give correct results for the Ising-ASFL model,
even for this small system size.
}\label{paper1_problem_of_2level}
\end{figure}

However, this apparently reasonable approach does not yield the correct DOS for the
two-dimensional Ising-ASFL model.
The joint DOS $g(M,M_{s})$ should be independent of $H$ and $A$. Figures \ref{paper1_problem_of_2level} (b) to (d) show that the results obtained by this method
change significantly when $H$ and $A$ change, and all of them are quite different from the result obtained by exact combinatorial calculation in (a).

The reason for the differences can be explained as follows. Consider applying the two-stage method to get the DOS in terms of only two macroscopic variables, $(E,V_{1})$. In the second stage of the process, suppose a walker is trying to move from a phase point $(E,V_{a})$ to a point $(E,V_{b})$, and then to a point $(E,V_{c})$. As these three points have the same energy, the moves will have the same acceptance probabilities, $1/g(E)$. In this sense, these moves will be similar
to an unbiased random walk confined to an energy $E$, and with the histogram corresponding to $V_{1}$ being recorded.
Thus, if the change in $g(E,V_{1})$ along $V_{1}$ for a given value of $E$ is not too large,
the method should give good results. However, if $g(E,V_{1})$ changes significantly when $V_{1}$ changes, the method may not give correct statistics.
Unbiased sampling works for extremely small systems where the differences in $g(E)$ are small,
but the WL method is required if the differences are significant.

The method can in principle be improved by doing separate WL simulation for each energy $E$ to find the statistics corresponding to $V_{1}$, if ergodicity is not broken. This observation leads to the basic principle of our macroscopically constrained WL method. To
preserve ergodicity and obtain $g(E,M,M_{s})$ for our system, we can simplify the process to just
perform simulations in the {\it energy space\/} for fixed $M$ and $M_{s}$. We discuss the method in detail in the following sections.

\subsection{Other methods}\label{sec_others}

A different WL method to obtain $g(E,M)$ was proposed by  Zhou \textit{et al.} \cite{PhysRevLett.96.120201}. A kernel function is applied when one tries to update the histogram.
The method appears to save time, but it is quite complicated and tuning of kernel functions seems to be
required for different systems.

Very recently, Zablotskiy \textit{et al.} \cite{PhysRevE.93.063303} used the stochastic approximation
MC method to obtain the joint DOS, $g(V_{2},V_{1})$, of a polymer model, and then used it to deduce the $g(E)$ of the system. The way they obtain $g(V_{2},V_{1})$ is  similar to the method in Sec.~\ref{sec_2D_WL} but including the $1/t$ algorithm, and only small $\ln g(V_{1},V_{2})$ were considered.

Two papers have recently been published that use methods similar to, but less general to the
one presented here. Louren\c{c}o and Dickman \cite{1742-5468-2016-3-033107}
obtained the two-dimensional joint DOS for the square-lattice
Ising antiferromagnet, $g(E,M)$ at a single phase point, using the tomographic entropic sampling method \cite{PhysRevE.84.026701,1742-5468-2014-7-P07007}.
The method employs many random walkers, each starting in a different energy state, and their results are combined  to get $g(E,M)$.
From this, they obtained the critical line and canonical averages of the order
parameter, $M_s$.
However, as pointed out in \cite{PhysRevE.84.026701}, the tomographic entropic sampling method cannot give correct results when the system size is large.
In a study motivated by a network reliability problem,
Ren, Eubank, and Nath presented a method
to obtain the joint DOS for the square-lattice Ising ferromagnet, $g(E,M)$, using parallel
WL simulations at fixed $M$ \cite{PhysRevE.94.042125}. While conceptually similar to the method we present here, we
note that they do not discuss generalizations to higher-dimensional order-parameter
spaces. Moreover, we will here discuss several methods to simplify the computational process.

\section{Macroscopically constrained Wang-Landau }\label{sec:our_WL}
\subsection{Basic idea}\label{sec:our_idea}
 Suppose one wants to obtain the joint DOS for a system with $K$ macroscopic variables, $g(V_{K},...,V_{2},V_{1})$. Instead of letting the random walker travel in a $K-$dimensional phase space ($V_{K}$,...,$V_{2}$,$V_{1}$), which would require a very long time to obtain a `flat' histogram, the simulation can be broken into many simulations performed in smaller phase spaces as follows.
 First, obtain $g(V_{1})$ through normal WL simulation or direct calculation. Then, break the large phase space into smaller phase spaces, each with a different fixed value of $V_{1}$.
  For each value of $V_{1}$, a separate WL simulation is performed to obtain the
DOS with respect to only one macroscopic variable $V_{2}$, denoted as $g(V_{2}|V_{1})$. Next, each phase space is broken into smaller phase spaces, each with a different fixed value of $(V_{2},V_{1})$. Again, separate simulations for different fixed values of $(V_{2},V_{1})$ are performed to obtain DOS with respect to only one macroscopic variable $V_{3}$, denoted as $g(V_{3}|V_{2},V_{1})$.
Iterating the process, the joint DOS with $K$ variables, $g(V_{K},...,V_{2},V_{1})$, can be obtained as
 \begin{eqnarray}
 g(V_{2},V_{1})&=&\frac{g(V_{2}|V_{1})}{\sum_{V_{2}}g(V_{2}|V_{1})}g(V_{1})     \label{def_joint_DOS_principle1} \\
g(V_{3},V_{2},V_{1})&=&\frac{g(V_{3}|V_{2},V_{1})}{\sum_{V_{3}}g(V_{3}|V_{2},V_{1})}g(V_{2},V_{1})    \label{def_joint_DOS_principle2} \\
\vdots \nonumber \\
 g(V_{K},...,V_{1})&=&\frac{g(V_{K}|V_{K-1},...,V_{1})}{\sum_{V_{K}}g(V_{K}|V_{K-1},...,V_{1})}g(V_{K-1},...,V_{1}).
 \label{def_joint_DOS_principle10}
\end{eqnarray}
 In general, the macroscopic variables should be arranged such that the more fundamental building blocks of $g(V_{K},...,V_{1})$, like $g(V_{1})$, $g(V_{2},V_{1})$ and $g(V_{3},V_{2},V_{1})$, can be obtained in the most accurate manner.
Therefore,
if the joint DOS for two macroscopic variables 
can be obtained directly by exact calculation, they should be chosen
as $V_{1}$ and $V_{2}$, so that the joint DOS $g(V_{K},...,V_{1})$ involves an exact factor, $g(V_{2},V_{1})$.  However, when partitioning the simulations into different stages, one
must be careful that in each stage, a simple method can be found to let the walker walk
through the whole confined phase space, such that ergodicity is not broken.

Each time the walker performs a WL process with only one free macroscopic variable, the constrained DOS, e.g., $g(V_{4}|V_{3},V_{2},V_{1})$, can be partitioned into different windows of $V_{4}$, and then joined together smoothly through choosing the contact point with the most similar slopes with respect to $V_{4}$ as in REWL \cite{PhysRevLett.110.210603,WaiReplica}.

Breaking down the single WL processes into many independent processes like this works fast and
is more accurate compared to the methods discussed in Sec.~\ref{sec:current_methods}. Furthermore, the algorithm itself is very suitable for parallelization on many independent processors.

\subsection{$g(E,M,M_{s})$ for the Ising-ASFL model}\label{sec:apply_on_Ising_system}

To obtain the joint DOS, $g(E,M,M_{s})$, for the Ising-ASFL model, we choose $V_{1}=M_{s}$, $V_{2}=M$, and $V_{3}=E$. This is convenient
because we can  calculate $g(M,M_{s})$ exactly through direct combinatorial
calculation as shown in Appendix A.
Therefore, we can
 directly arrive at Eq.~(\ref{def_joint_DOS_principle2}) and write
\begin{equation}\label{def_joint_DOS}
g(E,M,M_{s})=\frac{g(E|M,M_{s})}{\sum_{E}g(E|M,M_{s})}g(M,M_{s}).
\end{equation}
Separate independent WL
simulations will be performed for different fixed values of $(M,M_{s})$, each obtaining a DOS
in terms of one macroscopic variable ($E$), $g(E|M,M_{s})$.

Any square lattice can be simply broken down into two sublattices, $A$ and $B$.
Every alternate site belongs to the same sublattice. 
The magnetization ($M$) and the staggered magnetization ($M_{s}$) can be written in terms of the magnetizations of these two sublattice, $M_{A}$ and $M_{B}$, as
\begin{eqnarray}
 M &=& M_{A}+M_{B}     \label{def_mag_us} \\
 M_{s}  &=& M_{A}-M_{B}  ~.  \label{def_stgmag_us}
\end{eqnarray}
Exchanging spins (Kawasaki dynamics) independently on each
sublattice will preserve $M_{A}$ and $M_{B}$, and thus also preserves the values of $M$
and $M_{s}$. Moreover, it will allow the walker to walk through all the possible configurations
and energies corresponding to each ($M,M_{s}$),
and thus preserve ergodicity. This is the method used here
to perform the random walk in microstates.

\subsection{Advantages of obtaining $g(E,M,M_{s})$}\label{sec:advantage_shifting}
If we want to know the DOS $g(E)$ under different conditions, say for different external magnetic fields $H$,
 and different long-range interaction strengths $A$,
  using the simple WL method or importance sampling MC, we would have to
perform separate runs every time these conditions are changed. However,
if we can obtain $g(E,M,M_{s})$, we only have to do WL for a
single set of $H$ and $A$.
For simplicity we choose zero field and zero long-range interaction, i.e., $H=A=0$.
The results for other parameter values
can be obtained by simply shifting the result obtained for $H=A=0$.
This happens because all the microstates are equally shifted in energy when a
field-like model parameter changes, as the field is coupled to a global property, such as $M$, according to Eq.~(\ref{def_Hamiltonian_2D_Ising-ASFL}).
Therefore, we can shift the DOS result from $H=A=0$ to the DOS
for arbitrary $H$ and $A$ through the transformation,
 \begin{equation}\label{def_shift_E_DOS}
g(E,M,M_{s}) \rightarrow g(E-HM - \frac{AM^{2}}{2N},M,M_{s}).
\end{equation}
This shifting approach saves a very large amount of work.

\subsection{Simplification through symmetry considerations}\label{sec:simplication_symmetry}
\begin{figure}
\includegraphics[width=0.5\textwidth]{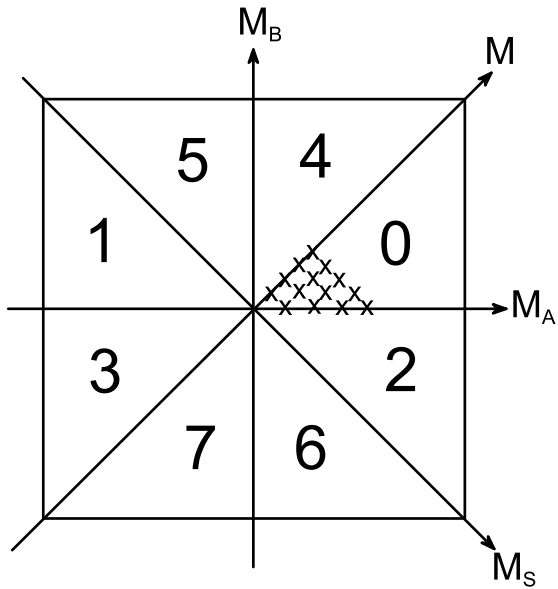}
\caption{
Uniform sampling of ($M$,$M_{s}$) pairs is performed in region 0. Some data points are illustrated as black crosses in the figure. After the sampling in region 0 is finished, symmetries are
used to obtain data for all the other seven regions.
}\label{WL_fig_symmtry_sq}
\end{figure}

Through the use of the shifting approach described in Sec.~\ref{sec:advantage_shifting}, we only have to consider $H=A=0$. 
This enables further simplification through symmetry considerations.
Consider a spin configuration (microstate) that belongs to the macrostate $(M,M_{s})$ lying in region 0 of Fig.~\ref{WL_fig_symmtry_sq}. For $H=A=0$, if we flip all the spins on sublattice $A$, $E$ and $M_{A}$ of the system will be reversed. From Eqs.~(\ref{def_mag_us}) and (\ref{def_stgmag_us}),
$M$ and $M_{s}$ of this new microstate are related to the original $M_{A}$ and $M_{B}$ through
\begin{eqnarray}\label{def:sym1_mag_magstg1_0}
 M &=& -M_{A} + M_{B}      \\
 M_{s} &=& -M_{A} - M_{B} \ .
\end{eqnarray}
Therefore, we have
\begin{equation}\label{def_dos_sym1_0}
g(E,M,M_{s}) = g(-E,-M_{A}+M_{B},-M_{A}-M_{B}) \ .
\end{equation}
This means that if we have obtained $g(E,M,M_{s})$ at one sampling point $(M,M_{s})$ in region 0, we can directly obtain $g(E,M,M_{s})$ at another point in region 1.
There are seven similar symmetry considerations, which correspond to regions 1-7 in Fig.~\ref{WL_fig_symmtry_sq}.
It is important to note that $g(E,M,M_{s})$ must not be double-counted along the four symmetry axes in Fig.~\ref{WL_fig_symmtry_sq} when combining the results.
These symmetry considerations reduce the computational work by roughly a factor of eight.
In Appendix B, we show explicitly how to map from region 0 to the other seven regions.

\subsection{Simplification through uniform sampling}\label{sec:simplication_sampling}
In region 0 of Fig.~\ref{WL_fig_symmtry_sq}, after we have chosen a data point at ($M,M_{s}$)=($0,0$), the next possible pairs are ($0,2$) and ($2,0$), i.e. the smallest increment is
$M_{\rm const}=2$. For a big system, the number of possible $(M,M_{s})$ pairs is very large. To get the DOS for every pair of $(M,M_{s})$ would require a huge amount of computational resources. Therefore,
data points in the $(M,M_{s})$ space are chosen with a constant increment, $M_{\rm const}$.
Here, choosing $M_{\rm const}=32$ gives good results for a system size of $L=32$.
Proper values for $M_{\rm const}$ for different system sizes will be further discussed in
Sec.~\ref{sec:generalizion big scaling}.

\subsection{Simulations in practice}\label{sec:practice}

\begin{figure}
\includegraphics[width=0.5\textwidth]{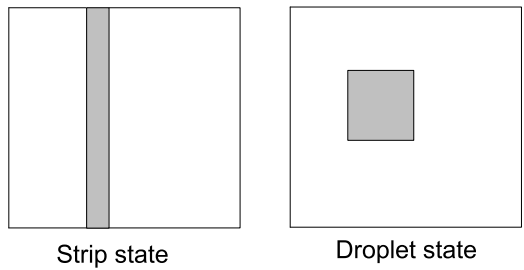}
\caption{Energy states that are close to the extreme energy states. The spins are aligned close to either a strip or a droplet shape \cite{0305-4470-23-20-021}.
}\label{WLfig_strip_droplet}
\end{figure}

With the simplifications introduced in the previous sections, WL processes can be carried out for separate ($M,M_{s}$). Here, we just list a few points  adopted in our simulations. The detailed implementation is described in Appendix C.

First, every ($M,M_{s}$) pair has a different accessible range of energies, and it is known that the extreme energy states have spins aligned close to either a strip or a droplet shape \cite{0305-4470-23-20-021} (Fig.~\ref{WLfig_strip_droplet}). We first estimate the extreme energies and then decide how many energy windows shall be used for each ($M,M_{s}$) pair.

Second, starting from the extreme energy states, artificial spin exchange processes are carried out in the initialization stage, so that we can find  more accessible energy states at the beginning of the simulation.

Third, the root-mean-square `flatness' criterion \cite{BROWN201428} is used, such that a histogram is regarded as `flat' according to  the root-mean-square deviation criterion,
\begin{equation}\label{def_release_deviation}
\sqrt{\frac{\sum_{E} |H(E) - H_{\rm average}|^{2}}{N_{E} }} < 20\% \ ,
\end{equation}
where $N_{E}$ is the number of accessible energy levels in that energy window, and the summation runs over these energy levels. This relaxation of the `flatness' criterion
can  make simulations finish much earlier, as already demonstrated in Sec.~\ref{sec_2D_WL}.

Fourth, the small statistical fluctuations in the DOS found at $H=A=0$
may be magnified near a critical point, causing difficulties in locating it accurately.
Here we reduce the statistical fluctuations by obtaining 10 different $g(E,M,M_{s})$ through
independent simulations, and taking the ensemble average.

\subsection{Simulation time}\label{sec:simulation_time}
For $L=32$, which is the largest system size we have considered, we have kept around 125 energy levels in edge windows and around 200 energy levels in non-edge windows. Most non-edge windows can finish simulations in a few minutes, but the edge windows, which contain energy levels with low density of states,
may take 20 min to more than one hour to finish. Therefore, most pairs of ($M,M_{s}$) considered can finish the simulation within a few minutes to a few hours. However, some edge windows may get `stuck' at energy levels with low DOS \cite{Yin20121568,PhysRevE.88.053302,PhysRevE.92.023306} and do not converge after several days, especially when the sampling points include $M_{s}=0$. Therefore, we reject a simulation that does not finish in two days and re-start the run. Some sampling points may have to be rejected and re-started several times.
With around two hundred computing cores, all the simulations for the $L=32$ system (including data for 10 different ensembles) could be finished within one week, with most of the time devoted to these `stuck' sampling points. If one intends to take the ensemble average of 10 different $g(E)$ as we do here, one may submit more than 10 identical jobs for the sampling points with $M_{s}=0$ at the beginning, and reject all the runs for the remaining jobs after 10 of them have finished. This can make the simulation finish earlier. The `stuck' problem is further discussed
in Sec.~\ref{sec:generalizion big problem}.

\section{Application of $g(E,M,M_{s})$}\label{sec:application}
\subsection{Density of states for arbitrary $H$ and $A$}\label{sec:density_states}

\begin{figure}
\includegraphics[width=0.45\textwidth]{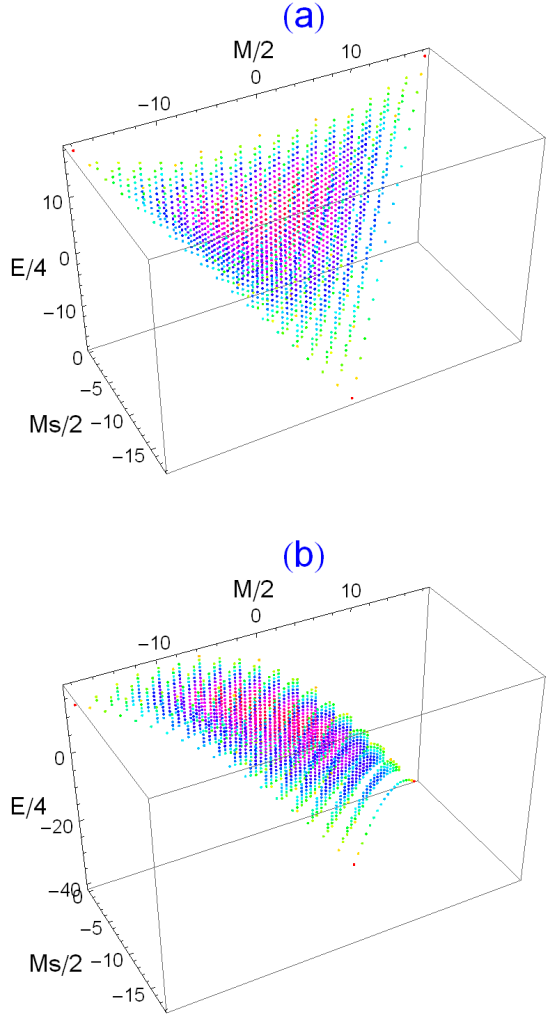}
\caption{
(Color online) $\ln g(E,M,M_{s})$ vs $E/4$, $M/2$ and $M_{s}/2$ for (a) $(H,A)=(0,0)$, and (b) $(H,A)=(3,7)$, both using $L=6$ for improved visibility. The results for larger systems are
similar.
$g(E,M,M_{s})$ at $(H,A)=(3,7)$  is obtained by shifting the $g(E,M,M_{s})$ at $H=A=0$
through Eq.~(\ref{def_shift_E_DOS}).
 The color of the data points shows the relative magnitude of the natural logarithm of the
DOS, ranging from red (smallest) to magenta (largest).
Only results for $M_{s}\leq 0$ are shown as there is reflection symmetry about the $M_{s} = 0$ plane.
}\label{paper1_gEMMs_H0_H3}
\end{figure}

\begin{figure}
\includegraphics[width=0.5\textwidth]{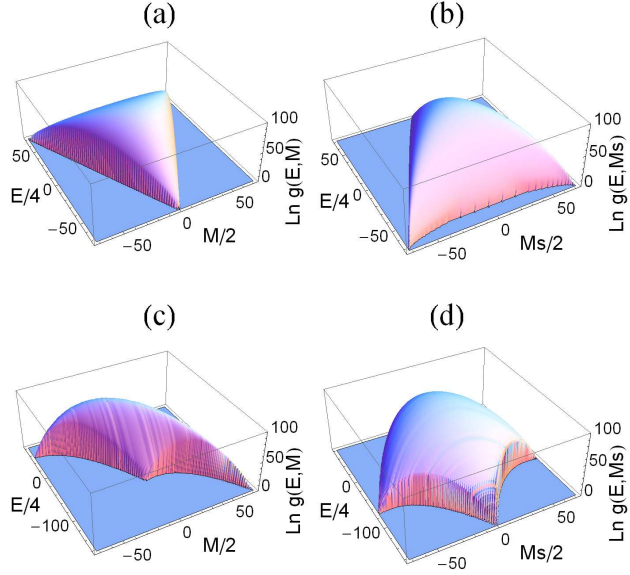}
\caption{
(Color online) $\ln g(E,M)$ vs $E/4$ and $M/2$ for (a) $(H,A)=(0,0)$, and (c) $(H,A)=(3,7)$.  $\ln g(E,M_{s})$ vs $E/4$ and $M_{s}/2$ for (b) $(H,A)=(0,0)$, and
(d) $(H,A)=(3,7)$. All using $L=12$. All the data are obtained by summing over the contribution of different directions in $g(E,M,M_{s})$. $g(E,M,M_{s})$ at $(H,A)=(3,7)$ are obtained by shifting the $g(E,M,M_{s})$ at $H=A=0$ through Eq.~(\ref{def_shift_E_DOS}), i.e., all data is obtained from   $g(E,M,M_{s})$ at $H=A=0$.
}\label{paper1_gEMgEMs_H0_H3}
\end{figure}

Figure \ref{paper1_gEMMs_H0_H3}(a) shows the joint
DOS $g(E,M,M_{s})$ for $H=A=0$ obtained from our simulation for the Ising-ASFL model (Sec.~\ref{sec:Hamiltonian_Ising_ASFL}). By shifting the energy as stated in Sec.~\ref{sec:advantage_shifting}, we obtain $g(E,M,M_{s})$ for $(H,A)=(3,7)$ as shown in Fig.~\ref{paper1_gEMMs_H0_H3}(b). Figure \ref{paper1_gEMgEMs_H0_H3} shows that by summing over one component of $g(E,M,M_{s})$, we can obtain $g(E,M)$ and $g(E,M_{s})$, which give very smooth results. Indeed, $g(E,M,M_{s})$, $g(E,M)$, $g(E,M_{s})$ and $g(E)$ for arbitrary $H$ and $A$ can be obtained in this way.

\subsection{Drawing phase diagrams}\label{sec:maths_phase_diagram}

We introduce the normalized magnetization and staggered magnetization as $m=M/L^{2}$ and $m_{s}=M_{s}/L^{2}$ respectively, which are the order parameters of our system.
After obtaining $g(E,M,M_{s})$ at $H=A=0$ through the simulations, we can use Eq.~(\ref{def_shift_E_DOS}) to get $g(E,M,M_{s})$ at arbitrary points in the
phase diagram and calculate different quantities as follows.

\begin{figure}
\includegraphics[width=0.5\textwidth]{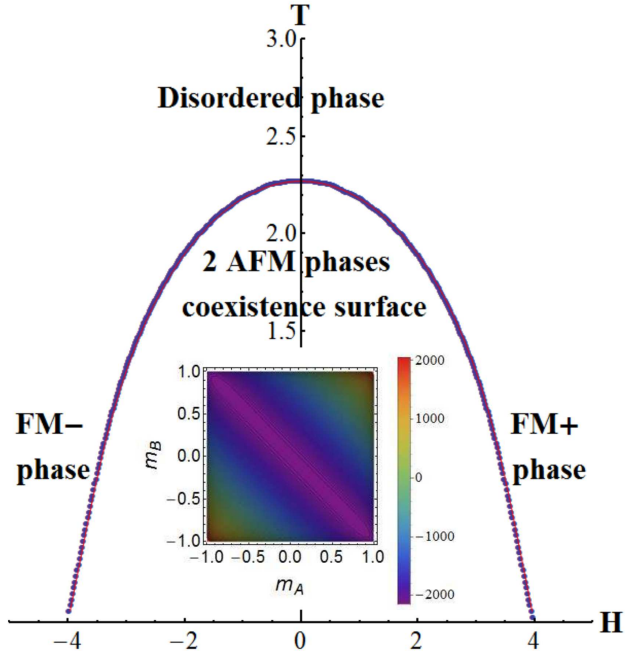}
\caption{(Color online) Critical line for an antiferromagnetic Ising system. The blue critical line is obtained by increasing $H$ from 0 in steps of  0.01 or 0.02, then performing a temperature ($T$) scan, choosing $\Delta T$ to be 0.001 to 0.005, and locating the critical line by choosing the point that gives the cumulant value (Eq.~(\ref{def_cumulant}) with $p=m_{s}$) closest to 0.61 \cite{0305-4470-26-2-009}. Data points on the negative $H$ side are obtained by reflection. The analytically approximated critical line (red), obtained by the method of Ref.\ \cite{WU1990123},
is also plotted. The two results coincide at this resolution.
The inset shows the free-energy contour diagram at the critical temperature for $H=0$ (refer to Eq.~(\ref{def:Free_energy_pt})).
}\label{paper1_Ising_phase}
\end{figure}

We can define the constrained partition function of any macrostate $(m,m_{s})$ as
\begin{equation}\label{def:partition function_pt}
Z_{m,m_{s}}=\sum_{E} g(E,m,m_{s}) e^{-E/T} ~.
\end{equation}
The overall partition function of the system is then
\begin{equation}\label{def:partition function_all}
Z_{\rm{all}}=\sum_{m,m_{s}} Z_{m,m_{s}} ~.
\end{equation}
The joint probability of finding the system in a macrostate $(m,m_{s})$ is
\begin{equation}\label{def:joint_Prob_density}
P(m,m_{s})\Delta m \Delta m_{s}=\frac{Z_{m,m_{s}}}{Z_{\rm{all}}} ~,
\end{equation}
where $\Delta m$, $\Delta m_{s}$ are the step sizes, both chosen to be the same value, $M_{\rm{const}}/L^{2}$.
The free energy of macrostate $(m,m_{s})$ is
\begin{equation}\label{def:Free_energy_pt}
F(m,m_{s})= -T \ln Z_{m,m_{s}} ~.
\end{equation}
As $(m_{A},m_{B})$ has a one-to-one relation with $(m,m_{s})$, we may express these quantities in terms of $(m_{A},m_{B})$, as well. The inset in Fig.~\ref{paper1_Ising_phase} shows a free-energy contour diagram $F(m_{A},m_{B})$ close to the critical temperature for $H=A=0$.

We can sum over the contributions of the joint probability (Eq.~\ref{def:joint_Prob_density}) in one direction,
obtaining the marginal probability density as
\begin{eqnarray}
 P(m)\Delta m &=& \frac{\sum_{m_{s}} Z_{m,m_{s}}}{Z_{\rm{all}}}      \label{def:marginal_prob_m} \\
 P(m_{s})\Delta m_{s}  &=& \frac{\sum_{m}  Z_{m,m_{s}}}{Z_{\rm{all}}}  ~.  \label{def:marginal_prob_ms}
\end{eqnarray}
With these distributions, we can calculate the expectation values of the order parameters and other quantities. In a complicated phase diagram that involves metastable phase regions, the stable phase will be the phase that has the larger total area in the marginal probability density, rather than the phase that shows the higher peak. 

To locate and classify the critical points or lines between ordered and disordered phases in phase diagrams,
the fourth-order Binder cumulant is often used \cite{Landau_simulation_book,PhysRevLett.47.693, PhysRevB.30.1477,PhysRevB.34.1841,PhysRevB.84.054433,PhysRevE.91.012103},
 \begin{equation}\label{def_cumulant}
u_{p}=1-\frac{\langle(p -\langle p \rangle)^{4}\rangle}{3\langle(p-\langle p \rangle)^{2}\rangle^{2}} ~,
\end{equation}
where $p$ is the order parameter of the system.
As an illustration, we consider $A=0$ in the Ising-ASFL model, which is just a purely antiferromagnetic Ising model. 
It is commonly accepted that this critical line is in the Ising universality class, which (assuming isotropy and periodic boundary conditions) has a cumulant value near $0.61$ \cite{0305-4470-26-2-009}.
Therefore, using Eqs.~(\ref{def:marginal_prob_ms}) and (\ref{def_cumulant}) with $p=m_{s}$,  we locate the critical line 
by finding the phase point with cumulant closest to $0.61$. The results are shown in
Fig.~\ref{paper1_Ising_phase}, using $L=32$.
The critical line obtained with our method is smooth, and the excellent agreement with the
Wu and Wu analytic approximation \cite{WU1990123} and the simulation results of
Louren\c{c}o and Dickman \cite{1742-5468-2016-3-033107} indicate that our procedure of
restarting `stuck' simulation runs (see Secs.\ \ref{sec:simulation_time}
and \ref{sec:generalizion big problem}) does not lead to significant numerical inaccuracies.

\begin{figure}
\includegraphics[width=0.5\textwidth]{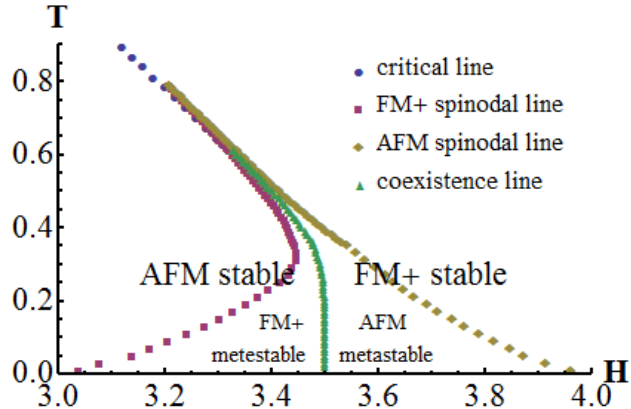}
\caption{(Color online)
The low-temperature portion of the phase diagram for $A=1$, near where the
critical line ends at a tricritical point,
using $L=32$. Metastable phases appear, leading to the appearance
of a coexistence line flanked by spinodal lines. The critical line is located by the
cumulant method (Eq.~(\ref{def_cumulant})) in the same way as in Fig.~\ref{paper1_Ising_phase} with $p=m_{s}$. The coexistence line is located by finding the maximum
susceptibility (Eq.~(\ref{def_susceptibility0})) with $p=m$ when $H$ is changed
at constant $T$. Spinodal lines are located by finding the value of $H$ where the free energy $F(m)$ (Eq.~(\ref{def:marginal_F_m})) changes from having two local minima to only one local minimum. The positions of the lines at $T=0$ agree with  exact ground-state calculations \cite{PhysRevB.93.064109}.
}\label{ASFL_phase_diagram_A1end}
\end{figure}

Another common boundary line that separates different phases in the phase diagram is a first-order phase transition (coexistence) line. We locate it by looking at the order-parameter variance, which is proportional to the susceptibility times the temperature,
\begin{equation}\label{def_susceptibility0}
var(p)=\chi_{p}T = L^{2}( \langle p^{2}\rangle - \langle p \rangle^{2} )  ~.
\end{equation}
This quantity has a local maximum value when evaluated at a point on the coexistence line, which serves as an accurate tool to determine this line.
The line that becomes straight vertical for low $T$ in Fig.~\ref{ASFL_phase_diagram_A1end} is
a coexistence line obtained by using Eqs.~(\ref{def:marginal_prob_m}) and (\ref{def_susceptibility0}), with $p=m$ in the Ising-ASFL model with $A=1$.

\begin{figure}
\includegraphics[width=0.5\textwidth]{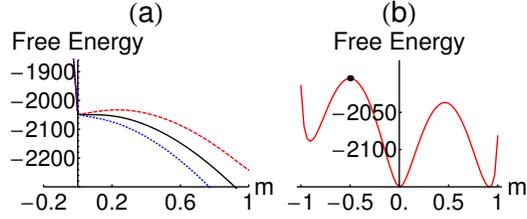}
\caption{(Color online) (a) Free energy $F(m)$ (Eq.~(\ref{def:marginal_F_m})) for $A=1$ and $T=0.11$, when the system is in the metastable region with two local minima at $H=3.689$ (red dashed),
 at the spinodal point at $H=3.789$ (black solid),
and  outside the metastable region at $H=3.889$ (blue dotted).
(b) Three different phases are shown in this free-energy diagram at $(H,T,A)=(2.267,0.033,8)$. The system is lying on the coexistence line between the phases in the middle and on the right-hand side. When locating the coexistence line through the susceptibility (Eq.~(\ref{def_susceptibility0})), one has to remove all the contributions from the metastable phase on the left-hand side of the black point.
}\label{paper1_spinodal_extra_phase}
\end{figure}

The metastable phase regions in this phase diagram are bounded by the spinodal lines.
We can express the free energy in terms of 
one order parameter as
\begin{eqnarray}
 F(m) &=& -T \ln \sum_{m_{s}} Z_{m,m_{s}}     \label{def:marginal_F_m} \\
 F(m_{s})  &=& -T \ln \sum_{m}  Z_{m,m_{s}} ~,  \label{def:marginal_F_ms}
\end{eqnarray}
and the spinodal lines  can be located by finding the points where the free energy changes from having two local minima to one local minimum. Figure \ref{ASFL_phase_diagram_A1end} illustrates two spinodal lines, and Fig.~\ref{paper1_spinodal_extra_phase}(a) shows the shape of the free energy at constant $T$ when the system is at and near a spinodal line.

It may happen that a coexistence line or a critical line between two different phases lies in the metastable regions of other phases. In those cases, we have to remove contributions from these non-relevant phases. Figure~\ref{paper1_spinodal_extra_phase}(b) illustrates one example.

\subsection{Probability densities and free energies at selected phase points}\label{sec:study_phase_diagram}

\begin{figure}
\includegraphics[width=0.48\textwidth]{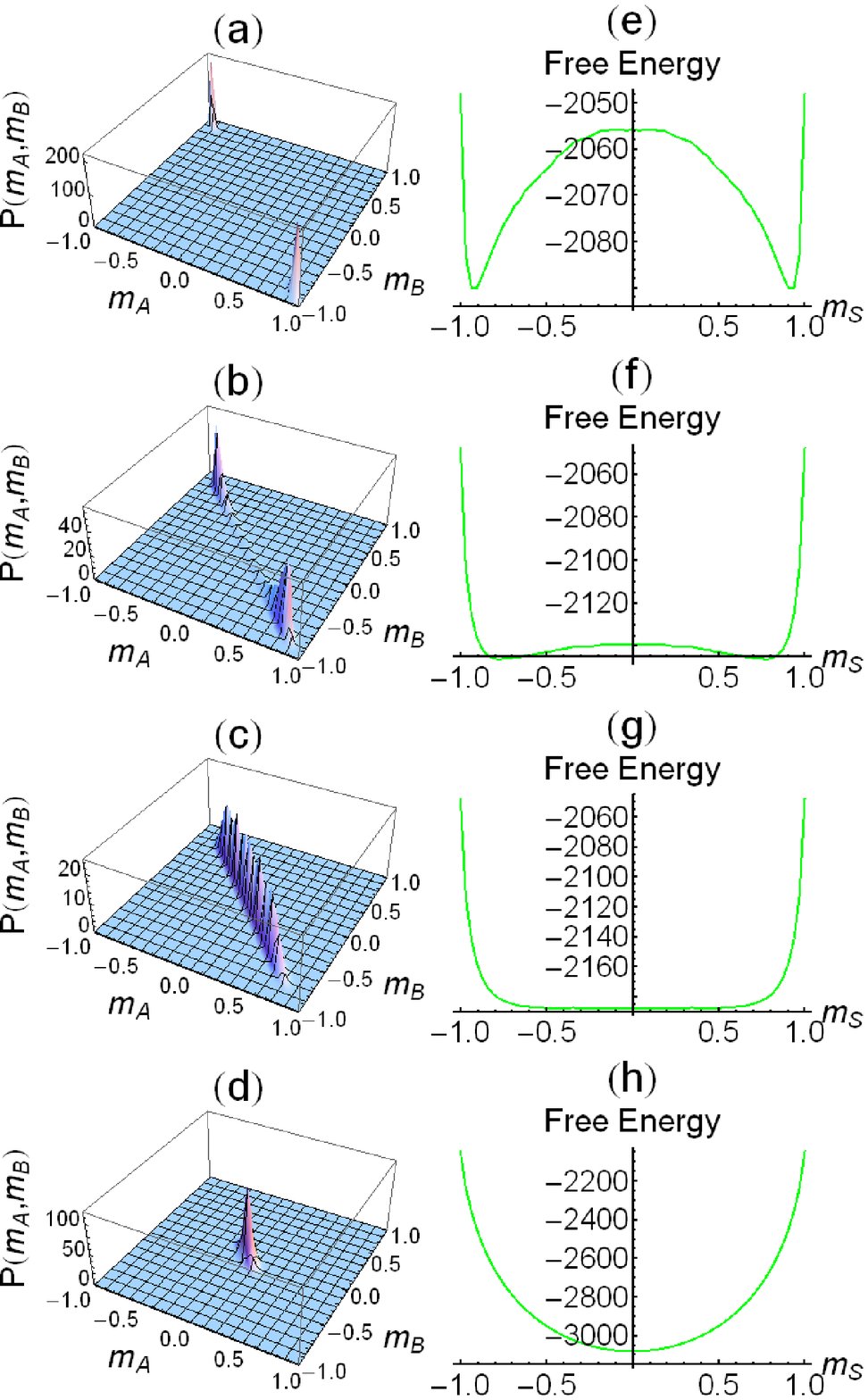}
\caption{(Color online) (a) - (d) Joint probability density $P(m_{A},m_{B})$ (Eq.~(\ref{def:joint_Prob_density}) with a change of variables), (e) - (h) corresponding free energy $F(m_{s})$ (Eq.~(\ref{def:marginal_F_ms})), when the system ($L=32$) is moved from a low temperature to a high temperature, crossing the critical line at $H=0$.
(a), (e)  $T=2<T_{c}$, where system is in one of the AFM phases with equal probability.  (b), (f) $T=2.275=T_{c}$ for this system size, where the two peaks are connected by a `bridge' and the cumulant is approximately equal to 0.61. (c), (g) $T=2.4>T_{c}$, where system is in the disordered phase with large AFM fluctuations. (d), (h) $T=4\gg T_{c}$, where the 
AFM fluctuations are much less pronounced. Note the different scales in the free-energy plots (e), (f), (g), (h).
}\label{paper1_crossing_cpline_T}
\end{figure}

\begin{figure}
\includegraphics[width=1\textwidth]{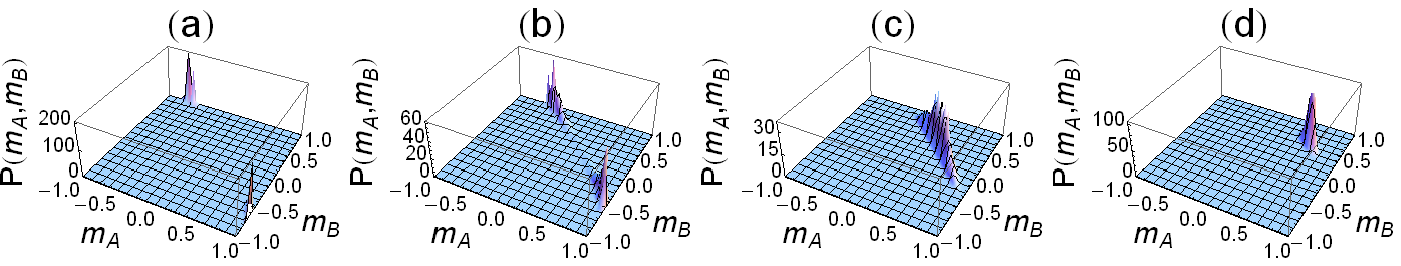}
\caption{(Color online) Joint probability density $P(m_{A},m_{B})$ (Eq.~(\ref{def:joint_Prob_density}) with a change of variables), when the system ($L=32$) is moved parallel to the $H$ axis at a low temperature, $T=0.06$, and crosses the critical line. (a) shows $H=3.95<H_{c}$, where system is close to the critical line but still in the AFM phases.  (b) shows $H=3.96=H_{c}$ for this system size and temperature, where the two peaks are connected by a `bridge' and the cumulant is approximately equal to 0.61. (c) shows $H=3.97>H_{c}$, where system is in the FM$+$ phase. (d) shows $H=4>H_{c}, $ where the FM$+$ phase 
peak is more symmetric and closer to the corner at  $m=1$.
}\label{paper1_crossing_cpline_H}
\end{figure}

As we can shift $g(E,M,M_{s})$ to obtain different quantities for any point in the phase diagram, we can analyze the free energy and probability density at any selected phase point in detail.

Figure \ref{paper1_crossing_cpline_T} shows what happens when moving along the temperature axis at $H=0$ and crossing the critical line in the antiferromagnetic Ising model, i.e., $A=0$. At low $T$, the joint probability density $P(m_{A},m_{B})$  has peaks only in the two antiferromagnetic (AFM) phases in the two opposite corners of the $(m_{A},m_{B})$ plane, $m_{s}=\pm 1$. Then, the two peaks connect weakly at the critical temperature. Above the critical temperature, the two peaks join, corresponding to a disordered phase. The free-energy contour diagram for the critical point at $H=0$ is shown as an inset in Fig.~\ref{paper1_Ising_phase}.

Similarly, Fig.~\ref{paper1_crossing_cpline_H}
demonstrates what happens when $H$ is increased  at constant, low $T$ to cross the critical line. The joint probability density $P(m_{A},m_{B})$ changes from two AFM phase peaks when the system is inside the critical line, to two peaks weakly connected at the critical line, and to a single ferromagnetic peak when the system is outside the critical line. The ferromagnetic peak moves toward the $(m,m_{s})=(1,0)$ corner and becomes more symmetric as $H$ further increases.

\section{Generalizing the scheme to bigger systems}\label{sec:generalizion big system}
\subsection{Scaling effect in choosing $M_{\rm const}$}\label{sec:generalizion big scaling}

\begin{figure}
\includegraphics[width=0.5\textwidth]{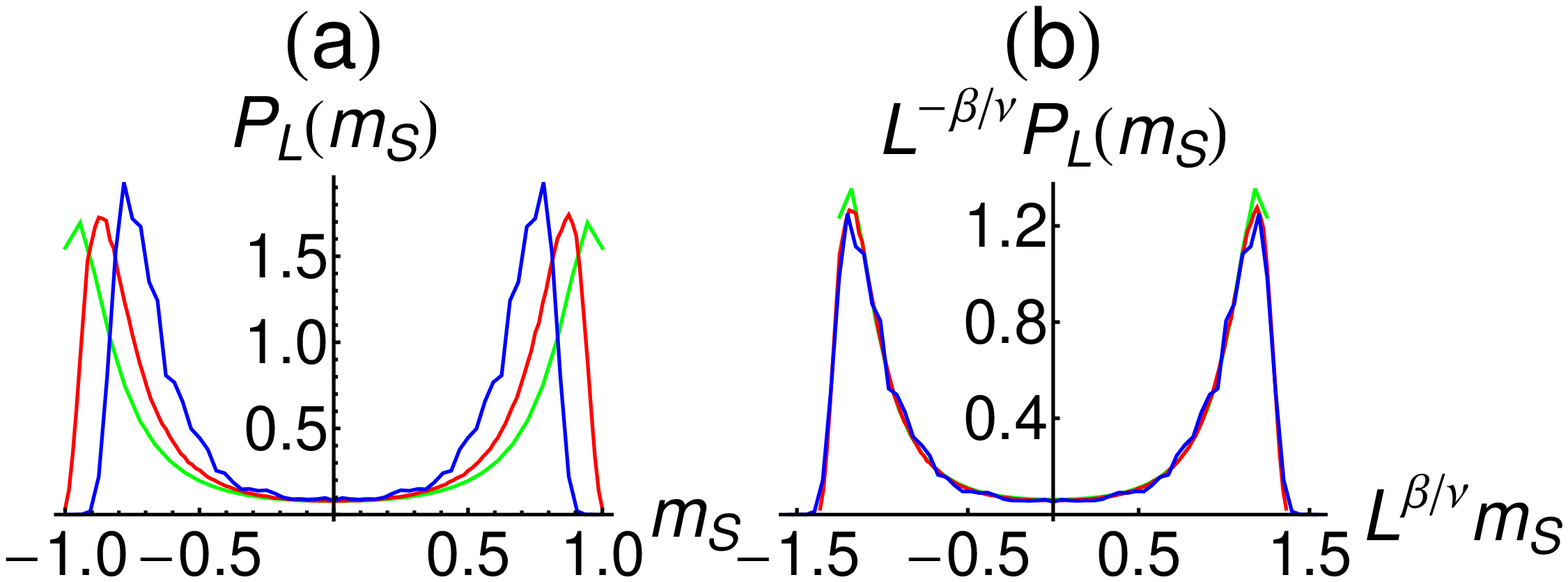}
\caption{(Color online) (a) Marginal probability density (Eq.~(\ref{def:marginal_prob_ms})) $P_{L}(m_{s})$ vs $m_{s}$ for $L=$ $6$ (green), $12$ (red), and $32$ (blue)  at $H=0$, and at their corresponding critical temperatures as determined from the cumulant value. (b) Shows the scaled plot, i.e., $L^{-\beta/\nu}P_{L}(m_{s})$ vs $L^{\beta/\nu}m_{s}$. The excellent data collapse indicates that even these small systems are fully in the asymptotic scaling regime, at least in this region of the phase diagram. It also indicates that any errors that might be caused by our
procedure of restarting `stuck' simulation runs (see Secs.\ \ref{sec:simulation_time}
and \ref{sec:generalizion big problem}) are insignificant.
}\label{ASFL_scaling_IS_critical}
\end{figure}

If the system size is large, in regions where phase transitions occur, the probability densities usually only show peaks in a small region of the order-parameter space.
Therefore, the resolution, $M_{\rm const}$, must be small enough to observe these peaks.
From finite-size scaling theory, plotting $L^{-\beta / \nu} P_{L}( m_{s} )$  vs $ L^{\beta / \nu}m_{s} $ for different system sizes $L$ will give curves that coincide (Fig.~\ref{ASFL_scaling_IS_critical}). That means that if we we can get a perfect result with a small system size $L_{s}$, when we consider a bigger system $L_{b}$,
the increment has to be chosen such that
$M_{{\rm const},b}\leq M_{{\rm const},s}\times (L_{b}/L_{s})^{2-\beta / \nu}$,
where $\beta$ and $\nu$ are critical exponents.
For critical points that belong to the Ising universality class,  $\beta / \nu = 1/8$,
whereas for the mean-field class, $\beta / \nu = 1/2$.

As the number of $(M,M_{s})$ pairs that must be sampled is
$L^{4}/M_{\rm const}^{2}$, the minimum number of $(M,M_{s})$ pairs required for
a big system is proportional to
$(L_{b}/L_{s})^{2 \beta / \nu}$. In this sense, as the Ising-ASFL model
contains mean-field critical points for large values of $A$ \cite{PhysRevB.93.064109},
when the system size is doubled, the number of $(M,M_{s})$ pairs must also be doubled.
If one is working with a purely short-range ferromagnetic or antiferromagnetic Ising model,
all critical points will be in the Ising class, so that the required number of $(M,M_{s})$
pairs one has to consider will be nearly unchanged ($\times 1.189$) if $L$ is doubled.

\subsection{Other problems for large systems, `flatness' criteria, and possible solutions}\label{sec:generalizion big problem}
The occurrence of `stuck' simulation runs has also been observed in other WL based
algorithms \cite{Yin20121568,PhysRevE.88.053302,PhysRevE.92.023306},
particularly in phase regions of extreme energy and/or very low DOS.
It is thus not a specific consequence of the macroscopic constraints in the present method.

In the work presented here, simulation runs for larger systems
sometimes get `stuck,' particularly at phase points with $M_{s} \approx 0$.
These points indeed correspond to extreme energy configurations arranged as perfect strips or
droplets (Fig.~\ref{WLfig_strip_droplet}). These states, or states that have similar energies,
are also very rare and thus have a very low $g(E|M,M_{s})$. Many of them are found in the initialization process, but they are very hard to reach during the subsequent simulation run.
For $L=32$, this problem can still be solved by rejecting and restarting the run without causing
significant sampling error. (See the numerical results in Secs.\ \ref{sec:application}
and \ref{sec:generalizion big scaling}.)
However, further increase in system size to $L=64$ causes more simulations to get `stuck.'

Using a `strict' flatness criterion as in  \cite{PhysRevLett.86.2050,PhysRevE.64.056101}, i.e.,
requiring the histogram of every energy level to not deviate too much from the average histogram, simulations may have great difficulty finishing. Although the presence of these extreme states is known through the initialization process, they are very hard to reach in the random walk process, so that the `flat' histogram may not be attainable. Significant improvement is obtained by
using the relaxed flatness criterion \cite{BROWN201428} as in Eq.~(\ref{def_release_deviation}),
by which the histogram is accepted as `flat,'
even though it deviates significantly from the average for these few energy levels.
Even without partitioning the energy spectrum into windows, runs that cannot finish in several
weeks with the strict flatness criterion can finish in a few days with the relaxed criterion.

However, this method also brings another problem. If a few energy levels are very difficult to reach and are not reached within the first few times a `flat' histogram is accepted, then, if one of these states is reached later, it will have $\ln g(E|M,M_{s})=0$ while other states that previously were visited can have $\ln g(E|M,M_{s})$ of the order of $10^{7}$. The large difference between the two neighboring states causes the walker to stay in this state to increase its $\ln g(E)$ until it reaches the order of $10^{7}$. However, every time a `flat' histogram is obtained, the increment, $\ln f$ is reduced by a factor of two, so that in this run, the increment step size may have already dropped to a very small value, like $10^{-6}$. Then it will take a very long time to raise $\ln g(E|M,M_{s})$ to the order of $10^{7}$, and the program will get stuck. As the difference between two neighboring states (or slope of $g(E|M,M_{s})$ vs $E$) increases with the system size,  a
simulation under these conditions may easily get `stuck' for large $L$.

To generalize the method to larger systems, one promising solution is dividing the edge
windows into smaller windows and applying REWL
\cite{PhysRevLett.110.210603,WaiReplica,PhysRevE.90.023302} in these windows, so that each walker is confined to a smaller energy range. This forces it to sample all these
energy levels with low density of states, and at the same time the replica can sample through all possible states in the edge windows.
Yet another way that may also alleviate the problem is to let the walker occasionally jump to
a previous state stored in a configuration database \cite{PhysRevLett.115.190602}.

\section{Conclusion}\label{sec:conclusion}
A macroscopically constrained WL method is proposed, which may be useful in
finding DOS with more than one variable, and in obtaining complex phase diagrams.
The method converts a multi-dimensional
random-walk process into many one-dimensional random walks, with each walker constrained to
fixed values of certain macroscopic order parameters.
The method is demonstrated and validated on
a two-dimensional antiferromagnetic Ising model with ferromagnetic long-range interaction.
We obtained the joint DOS $g(E,M,M_{s})$, through simulations at $H=A=0$. The
DOS for arbitrary values of $(H,A)$ then follow by a simple transformation
of the total system energy, and all the thermodynamic quantities for any point in the phase diagram can then be found. We demonstrate how to use the DOS obtained to efficiently draw
phase diagrams and free-energy landscapes for the Ising-ASFL model.

The detailed physics of the Ising-ASFL (spin-crossover material) model,
including complex phase diagrams  for several values of $A$,
will be described in  forthcoming papers \cite{CHAN17conference,CHAN17}.

\section*{ACKNOWLEDGMENTS}\label{sec:acknowledgement}

Chor-Hoi Chan thanks Alexandra Valentim and Ying-Wai Li for helpful discussions of the Wang-Landau and replica-exchange Wang-Landau methods. 
The simulations were performed at the Florida State University High Performance Computing Center
This work was supported in part by NSF Grant No. DMR-1104829.


\appendix
\section{}
\label{sec:exact_combinatorial}
The exact combinatorial calculation of $g(M,M_{s})$ for the Ising-AFSL model is described below.

Let $N_{A}$ be the number of sites on sublattice $A$ that have spin up, and $N_{B}$ be the number of sites on sublattice $B$ that have spin up. Then, the magnetization ($M$) and staggered magnetization ($M_{s}$) in Eqs.~(\ref{def_mag_us}) and (\ref{def_stgmag_us}) can be rewritten as
\begin{eqnarray}
 M &=& 2( N_{A} + N_{B} ) - N      \label{def:mag_stagmag_NaNb1} \\
 M_{s}  &=& 2 ( N_{A} - N_{B} )  .  \label{def:mag_stagmag_NaNb2}
\end{eqnarray}
As the joint DOS $g(M,M_{s})$ is defined as the total number of spin configurations (microstates) that have certain $(M,M_{s})$ values,
 $g(M,M_{s})$ can be visualized as the total number of ways to allocate $N_{A}$ upspins on sublattice $A$ and $N_{B}$ upspins on sublattice $B$. This can be expressed as the product of two binomial factors,
\begin{eqnarray}\label{def_combin_DOS}
 g(M,M_{s}) &=& C^{N/2}_{N_{A}} C^{N/2}_{N_{B}}       \\
   &=& \frac{(N/2)!}{N_{A}!(N/2-N_{A})!} \frac{(N/2)!}{N_{B}!(N/2-N_{B})!} ,
\end{eqnarray}
where $N/2=L\times L/2$ is the total number of sites on each sublattice. The general binomial recursive formula,
\begin{eqnarray}\label{def:combin_formula}
 C^{n}_{k} = C^{n-1}_{k-1} + C^{n-1}_{k} \ ,1\leq k \leq n-1
\end{eqnarray}
with boundary values
\begin{eqnarray}\label{def:combin_formula2}
 C^{n}_{0} = C^{n}_{n} = 1
\end{eqnarray}
is used to speed up the calculation. Overflow problems will be present if the value of $n!$ is too large, so Stirling's approximation is employed when $\ln g(M,M_{s})$ is greater than $700$:
\begin{eqnarray}\label{def:Stirling}
 \ln n! \approx n \ln n - n.
\end{eqnarray}

\section{}
\label{sec:symme}
The symmetry considerations used to map the DOS from region 0 to regions 1-7 in Fig.~\ref{WL_fig_symmtry_sq} are described below.

\textit{Region 1}: $M_{A}\rightarrow -M_{A}$ and $E\rightarrow -E$.

This is reflection about the $M_{B}$ axis, which means that,
if we change $M_{A}$ to $-M_{A}$ by flipping all the spins on sublattice A, the energy changes from $E$ to $-E$, and the new $M$ and $M_{s}$ are related to the original $M_{A}$ and $M_{B}$ through
\begin{eqnarray}\label{def:sym1_mag_magstg}
 M &\rightarrow& -M_{A} + M_{B}      \\
 M_{s} &\rightarrow& -M_{A} - M_{B} \ .
\end{eqnarray}
Thus,
 \begin{equation}\label{def_dos_sym1}
g(E,M,M_{s}) = g(-E,-M_{A}+M_{B},-M_{A}-M_{B}) \ .
\end{equation}

\textit{Region 2}: $M_{B}\rightarrow -M_{B}$ and $E\rightarrow -E$.

This is reflection about the $M_{A}$ axis, which means flipping all the spins on sublattice B. Thus,
 \begin{equation}\label{def_dos_sym2}
g(E,M,M_{s}) = g(-E,M_{A}-M_{B},M_{A}+M_{B}) \ .
\end{equation}

\textit{Region 3}: $M_{A}\rightarrow -M_{A}$ and $M_{B}\rightarrow -M_{B}$ and $E\rightarrow E$.

This is a combination of reflection about the $M_{A}$ axis and $M_{B}$ axis,
which means flipping all the spins on both sublattices. This preserves the energy, and we have
 \begin{equation}\label{def_dos_sym3}
g(E,M,M_{s}) = g(E, -M_{A}-M_{B},-M_{A}+M_{B}) \ .
\end{equation}

\textit{Region 4}: $M_{A}\leftrightarrow M_{B}$ and $E\rightarrow E$.

This is reflection about the $M$ axis, which means
exchanging all the spins between the two sublattices. This preserves the energy, and we have
 \begin{equation}\label{def_dos_sym4}
g(E,M,M_{s}) = g(E, M_{B}+M_{A},M_{B}-M_{A}) \ .
\end{equation}

\textit{Region 5}: $M_{A}\rightarrow M_{B}$ and $M_{B}\rightarrow -M_{A}$ and $E\rightarrow -E$.

This is a combination of reflection about both the $M_{s}$ and $M_{A}$ axes, which means
replacing spins on sublattice A by spins on sublattice B, and spins on sublattice B by the flipped spins on sublattice A. This reverses the energy,
 and we have
 \begin{equation}\label{def_dos_sym5}
g(E,M,M_{s}) = g(-E, M_{B}-M_{A},M_{B}+M_{A}) \ .
\end{equation}

\textit{Region 6}: $M_{A}\rightarrow -M_{B}$ and $M_{B}\rightarrow M_{A}$ and $E\rightarrow -E$.

This is a combination of reflection about the $M_{s}$ and $M_{B}$ axes, which means
replacing spins on sublattice A by flipped spins on sublattice B, and spins on sublattice B by the spins on sublattice A. This reverses the energy,
 and we have
 \begin{equation}\label{def_dos_sym6}
g(E,M,M_{s}) = g(-E, -M_{B}+M_{A},-M_{B}-M_{A}) \ .
\end{equation}

\textit{Region 7}: $M_{A}\rightarrow -M_{B}$ and $M_{B}\rightarrow -M_{A}$ and $E\rightarrow E$.

This is reflection about the $M_{s}$ axis, which means
replacing spins on sublattice A by flipped spins on sublattice B, and spins on sublattice B by the flipped spins on sublattice A. This preserves the energy,
 and we have
 \begin{equation}\label{def_dos_sym7}
g(E,M,M_{s}) = g(E, -M_{B}-M_{A},-M_{B}+M_{A}).
\end{equation}

\section{}
\label{sec:implementation}

Details of the implementation scheme for not too large systems are given below.

\textbf{Step 1}: Determine the desired combination of $M$ and $M_{s}$.
While $L=6$ and $12$ were simulated with all possible $(M,M_{s})$ pairs
(i.e., step size $M_{\rm const}=2$ ), $L=32$ was simulated with
$M_{\rm const}=32$, which gives good results.
The symmetries described in Sec.~\ref{sec:simplication_symmetry} and Appendix B ensure that we only have to choose data points within one octant of the $(M,M_{s})$ space.

\textbf{Step 2}: Each chosen pair of $(M,M_{s})$ is submitted as a parameter to identical, independent WL programs running on separate processing cores (or sequentially on one core). For each WL process, the values of $M$ and $M_{s}$ are conserved in every time step.  Note that fixing an ($M$,$M_{s}$) pair is the same as fixing an ($M_{A}$,$M_{B}$) pair (refer to Eqs.~(\ref{def_mag_us}) and (\ref{def_stgmag_us}) and Fig.~\ref{WL_fig_symmtry_sq}).

\textbf{Step 3}: Find the maximum and minimum energies of the system for each chosen $(M,M_{s})$ pair. For each $(M,M_{s})$ pair, divide the energy spectrum into windows if the number of energy levels is large. We keep around $200$ energy levels in each window (around $125$ energy levels in the edge windows), using $50\%$ overlap between neighboring windows.

For a fixed ($M$, $M_{s}$) pair, macrostates that have energies close to the extreme energies are in general either arranged in a configuration close to a strip or a nearly square droplet \cite{0305-4470-23-20-021} (Fig.~\ref{WLfig_strip_droplet}). We first prepare all the spins pointing up (purely ferromagnetic state
), and then flip the spins separately on the two sublattices sequentially one by one until we get a configuration (microstate) that satisfies the chosen ($M_{A}$,$M_{B}$) pair. This locates the highest energy microstate that is arranged close to a strip shape.
To locate the highest energy states that are close to a droplet form, we again prepare all the spins pointing up (purely ferromagnetic state), and calculate the approximate size of the droplet if a group of down-spins are arranged in a nearly droplet shape. Then we again flip the spins on each sublattice separately until $M_{A}$ and $M_{B}$ are satisfied. To locate the lowest energy states, we first prepare all spins arranged alternately up and down (purely antiferromagnetic
), and then we perform similar flips to get states close to antiferromagnetic strip and droplet shapes.

\textbf{Step 4}: List as many energy levels as possible that satisfy this ($M$,$M_{s}$) pair and initialize $\ln g(E|M,M_{s})=0$ for each energy level found. Using the microstates found in Step 3, which have near-extreme energies and already satisfy the ($M$,$M_{s}$) constraint, we randomly perform spin exchange separately on the two sublattices many times to arrive at most of the energy levels that satisfy the same $M_{A}$ and $M_{B}$.
For each energy window, we store one spin configuration as the initial configuration of that energy window.
If some energy levels are still missing in this step, they will be visited later during the random walk process. Each newly found state is immediately initialized to $\ln g(E|M,M_{s}) =0$.

\textbf{Step 5}: 
Randomly choose the energy window to start with, and load the spin configuration stored during the initialization stage to use as the starting configuration. Set the histogram $H(E|M,M_{s})=0$ for all the energies $E$ in that energy window.

\textbf{Step 6}:
Propose a move in microstates by
  doing spin-exchange (Kawasaki dynamics) between two sites with different spins on the same sublattice, and decide whether the move is accepted according to \begin{equation}\label{def_jump1}
p(E_{1} \rightarrow E_{2}) = \min[ \frac{g(E_{1})}{g(E_{2})}  ,1].
\end{equation}

\textbf{Step 7}: 
Update the DOS $g(E|M,M_{s})$ for each ($M$,$M_{s}$) pair as $\ln g(E|M,M_{s}) \rightarrow \ln g(E|M,M_{s}) + \ln f$ whenever the energy level $E$ is visited. At the same time, its histogram is updated as $H(E|M,M_{s}) \rightarrow H(E|M,M_{s}) + 1$

\textbf{Step 8}: 
Using the root-mean-square `flatness' criterion in Eq.~(\ref{def_release_deviation}),
check whether a `flat' histogram $H(E|M,M_{s})$ is obtained for each ($M$,$M_{s}$) pair after every $100\times L^{2}$ time steps, and reduce the modification factor $f$ to its square root value whenever a `flat' histogram is reached.

 When the code is found to be stuck and cannot finish after a sufficiently long time for a particular $(M,M_{s})$ pair, we reject that run and re-start the simulation for that $(M,M_{s})$ pair. This reject and rerun process may have to be performed a few times on one $(M,M_{s})$ pair in order to obtain a final converged result. Further discussion of the convergence is given in Secs.~\ref{sec:simulation_time} and \ref{sec:generalizion big problem}.

 $\ln g(E|M,M_{s})$ is shifted to a small value every time we get a `flat' histogram, so that we can avoid $\ln g(E|M,M_{s})$ accumulating to very large values that cause overflow problems.
Restart the run at one of the configurations that has nearly minimum histogram in the last run.
Repeat the random walk process until the modification factor is less than $10^{-8}$.

\textbf{Step 9}: Join the $\ln g(E|M,M_{s})$ curves obtained in different windows for each ($M$,$M_{s}$) pair at the energy where $\ln g(E|M,M_{s})$ in the two adjacent windows have the closest slopes with respect to $E$. Use formula (\ref{def_combin_DOS}) to obtain the exact number of microstates for each pair of ($M$,$M_{s}$).
 The overall DOS, $g(E,M,M_{s})$ or $\ln g(E,M,M_{s})$, can then be obtained for the whole system through Eq.~(\ref{def_joint_DOS}).

\textbf{Step 10}: The $g(E,M,M_{s})$ obtained still has numerical error.
To average out the error,
 10 different $g(E,M,M_{s})$ were obtained, and their ensemble average was used as $g(E,M,M_{s})$.





\begin{thebibliography}{75}
\bibitem{Endnote1}
If we break a square lattice into two sublattices, the magnetization on one sublattice minus the magnetization on the other sublattice is defined as the staggered magnetization, $M_{s}$. It gives extreme values when all the spins are aligned alternately up and down. It is  the antiferromagnetic  order parameter.
\end{thebibliography}


\ifx \manfnt \undefined \font\manfnt=logo10 \fi

\end{document}